# Temperature Dependent Zero-Field Splittings in Graphene

C. Bray[1,‡], K. Maussang[2,‡], C. Consejo[1], J. A. Delgado-Notario[3,4], S. Krishtopenko[1], I. Yahniuk[4,5], S. Gebert[1,2], S. Ruffenach[1], K. Dinar[1], E. Moench[5], J. Eroms[5], K. Indykiewicz[6], B. Jouault[1], J. Torres[2], Y. M. Meziani[3], W. Knap[4], A. Yurgens[7], S. D. Ganichev[4,5], F. Teppe[1*]

[1] L2C (UMR 5221), Université de Montpellier, CNRS, Montpellier, France

[2] IES (UMR 5214), Université de Montpellier, CNRS, Montpellier, France

[3] Nanotechnology Group, USAL-Nanolab, Universidad de Salamanca, Salamanca 37008, Spain

[4] CENTERA Laboratories, Institute of High Pressure Physics, Polish Academy of Sciences, 29/37 Sokołowska Str, Warsaw, Poland

[5] Terahertz Center, University of Regensburg, 93040, Regensburg, Germany

[6] Faculty of Electronics, Photonics and Microsystems, Wroclaw University of Science and Technology, 50-372 Wroclaw, Poland

[7] Department of Microtechnology and Nanoscience, Chalmers University of Technology, SE-412 96 Göteborg, Sweden

* frederic.teppe@umontpellier.fr

[‡] These authors have contributed equally to this work and share first authorship

## Abstract

Graphene is a quantum spin Hall insulator with a 45 μeV wide non-trivial topological gap induced by the intrinsic spin-orbit coupling. Even though this zero-field spin splitting is weak, it makes graphene an attractive candidate for applications in quantum technologies, given the resulting long spin relaxation time. On the other side, the staggered sub-lattice potential, resulting from the coupling of graphene with its boron nitride substrate, compensates intrinsic spin-orbit coupling and decreases the non-trivial topological gap, which may lead to the phase transition into trivial band insulator state. In this work, we present extensive experimental studies of the zero-field splittings in monolayer and bilayer graphene in a temperature range 2K-12K by means of sub-Terahertz photoconductivity-based electron spin resonance technique. Surprisingly, we observe a decrease of the spin splittings with increasing temperature. We discuss the origin of this phenomenon by considering possible physical mechanisms likely to induce a temperature dependence of the spin-orbit coupling. These include the difference in the expansion coefficients between the graphene and the boron nitride substrate or the metal contacts, the electron-phonon interactions, and the presence of a magnetic order at low

**temperature. Our experimental observation expands knowledge about the non-trivial topological gap in graphene.**

## Introduction

The intrinsic spin-orbit (ISO) coupling in graphene comes from carbon atom $d$ orbitals [1;2] which hybridize with the $p_z$ ones. Indeed, the $p_z$ orbitals themselves have no net orbital momentum along $z$, so the spin-orbit interaction is expected to be weak [3]. The intrinsic spin splitting is therefore proportional to the spin-orbit coupling of the $d$ states. The ISO opens a gap of opposite sign at each Dirac point of magnitude of about $\Delta_I = 45$ µeV [4;5;6]. Indeed, an electron state at $K$ valley and a hole state at $K'$ valley are connected by sublattice symmetry. The emergence of this gap therefore moves graphene from the family of Dirac semimetals to the one of quantum spin Hall insulators [7;8]. This topological phase results in the emergence of the edge states connecting electron and hole bands at different Dirac points [5]. However, this sublattice symmetry in graphene can be broken, for instance, by the coupling of graphene to the boron nitride substrate [9;10;11], which in its turn induces staggered potentials ($\Delta$ at A-sites and $-\Delta$ at B-sites) and thus open a bandgap of $2\Delta$. This makes it energetically favourable for the electrons to stay in one of the sub-lattices, resulting in pseudo-spin order [12]. This staggered sub-lattice potential competes with the ISO potential, as the former leads to a trivial band insulator state, while the latter induces a topological insulator phase. These phenomena of ISO coupling and spin splitting of sub-lattices were addressed both theoretically and experimentally on graphene on hBN [4, 5 and 6] by means of resistively detected electron spin resonance (ESR) at low temperatures. Indeed, when a magnetic field $B$ is applied perpendicularly to the graphene sheet, the Kramers pairs split into spin-up and -down states by the Zeeman energy, $g\mu_B B$. Measuring the spin-flip transitions by ESR techniques allows one to determine accurately the different zero-field splittings (ZFS) in graphene. From there, it becomes possible to study this ISO coupling as a function of various physical parameters, such as temperature, hydrostatic pressure, or different orientations between graphene layers. For instance, the ISO coupling has been studied recently by ESR technique in twisted graphene bilayer [13]. As the ISO coupling is a relativistic atomic phenomenon, one could think at first glance that it should not be affected by the lattice temperature. However, graphene's perfect crystal symmetry can be broken when suspended or placed under massive contacts. As it cools down, the contraction of the metal can indeed deform the thin layer of carbon atoms [14]. The thermally induced strain can therefore affect, on the one hand, the symmetry of the sublattice, and on the other hand, the mixing of different atomic orbitals, and as a result the zero-field splittings become temperature dependent. Note that the electron-phonon interaction actually induces most of the temperature dependencies of the electronic structure of semiconductors [15]. For instance, the coupling of spins to lattice vibrations in graphene, with emphasis on flexural modes, may have a great impact on the spin-orbit coupling [16]. Similarly, the Jahn-Teller effect can modify the orbital states in molecular systems through the Ham reduction factors [17], which can therefore change the energy of the resonant transitions of the system studied by ESR spectroscopy. Additionally, the possible presence of a magnetic order is also well known to modify the influence of the external magnetic field on the spin resonances. While some of these factors may be rather small, they are of paramount importance for understanding the details of the topological properties of in graphene.

In this work, we adapt from [18] a sub-THz magneto-photoconductivity technique to detect ESRs in monolayer and bilayer graphene embedded or not in hexagonal boron nitride (hBN). We measure several spin resonances and, by extrapolating their energy evolution to zero magnetic field, we observe two ZFS attributed to the sub-lattice and ISO potentials. The photoconductivity signal being greater there, we focused our study on the graphene samples encapsulated in hBN. Interestingly, the ZFS energies are very comparable in monolayer and bilayer graphene. By decreasing the temperature, an increase of the energies of the two ZFS is observed, which is discussed in the context of the thermal strain of the graphene layers, magnetic ordered phase, and the electron-phonon interaction.

## Results

Graphene-based Terahertz (THz) detectors are fast and sensitive devices, operating on different photoconductivity mechanisms in a wide range of frequencies [19,20]. These physical phenomena include bolometric [21], thermoelectric [22] and ratchet effects [23], as well as ballistic [24] and plasma waves effects [25]. In all these cases, the electric field of the incident THz wave is rectified and transformed into a potential drop between two contacts of the graphene-based sensor. This THz photoconductivity technique is sensitive to all kinds of conductivity changes in the material. In the presence of a perpendicular magnetic field, it is also sensitive to magnetic-field driven resistance oscillations [26], to the carrier's population imbalance between Landau levels, or to quantum fluctuations of conductance [27]. Therefore, a non-resonant (or broad band) signal in the photovoltage, $\Delta U$, is first expected when the resistance of the graphene sheet experiences Shubnikov-de Haas oscillations. But more interestingly, whatever the physical phenomena at the origin of the rectification, and the type of THz detector mentioned above, a resonant signal is also expected when the energy of the incident sub-THz radiation matches the Zeeman splitting of the electronic states. The photoconductive response is produced by a perturbation of the equilibrium electron spin polarization [28]. Our experimental method is indeed similar to the "Electrically Detected ESR" (EDESR) technique [29,30] which is very sensitive as it exploits the effects of spin-dependent interactions on the conductivity of the sample instead of directly measuring the sub-THz power absorbed by the spin system [31]. The particularity of our photoconductivity technique is based on the optimization of the electromagnetic coupling between the incident THz wave and the two-dimensional electron gas. This technique therefore represents an effective EDESR method capable of operating with a high sensitivity over a wide range of frequencies ranging from tens of GHz to THz (See Methods for more details). The origin of the EDESR signal in graphene has not been deeply discussed in previous works. The authors only mentioned the resonant microwave absorption at the spin-flip transition energy as the phenomenon responsible for the observed signal [5]. However, although not the focus of this paper, it is recalled that spin-to-charge conversion mechanisms, namely spin-dependent scattering, tunnelling, and recombination [32], are at the origin of this population change of the two Zeeman states inducing a variation in conductivity.

We carried out THz photoconductivity measurements in the presence of a perpendicular magnetic field with THz sensors based on two Ratchet devices on monolayer graphene (sample A) and bilayer graphene (sample B) both encapsulated in hBN, as well as on p-n junctions on CVD grown monolayer graphene (sample C). The manufacturing details of these devices and their mode of operation are explained in the "Methods" section. The experiments were carried using radiation frequencies between 45 GHz and 220 GHz (sweeping the magnetic field up to 9 T), and temperature range from 2.6 to 12 K. The amplitude of the detection signal and the signal-to-noise ratio were generally much greater in the two Ratchet devices than in the p-n junction. This is why we focus in the main text of this article on these ratchet sensors, although the first results on p-n junctions reported in the Supplementary Material section show the same trends and features in the results.

In both Ratchet sensors at low magnetic field, the photoconductivity signal shows usual Shubnikov-de Haas like oscillations at all frequencies, as observed elsewhere in different systems [33, 34]. With incident frequency of 112 GHz, at magnetic field B in the range from 3.5 T to 4.3 T, the spectra clearly exhibit three resonances whose positions are independent of the back gate voltage (Fig. 1 a). It should however be noted that the sign of the measured signal depends finely on various parameters and, in particular, on the gate voltage. It is therefore possible to have a peak on one series of curves and a deep on another although it is still the same resonance. Similar behaviour has already been noted in Ref. [15]. This is the reason why the data processing in the form of colour maps helps in the detailed analysis of the results. A black and white chart of the raw results is given as an example in Figure 1 b), in which it is possible to clearly distinguish the evolution of the various lines and oscillations as a function of the magnetic field and the gate voltage. The signal of interest is affected in the measured signal by a baseline that depends on the magnetic field and on the working frequency. In the case of Fig. 1b), this baseline is rather stable with gate voltage and resonances are clearly visible in the normalized initial signal. However, the signal baseline depends a lot on the probe frequency as a results of frequency dependence of the free-space wave to sample coupling. Furthermore, one expects amplitude of resonances to decrease with increasing temperatures, has a consequence of thermal distribution of excited states probed by the optical transition of the electromagnetic wave. In order to visualize more clearly the different optical transitions of interest, the first and second derivatives of the signal in regard to the magnetic field are plotted in red-blue colour maps as shown in Figures 1 c) and d). Indeed, assuming a baseline dependence in magnetic field smoother than resonances width, such a procedure should strongly reduce the relative amplitude of baseline and increase precision for peak localization (see Supp. Mat.). In the following work, all data were processed similarly, with peak localization performed on the second derivative signal to insure consistency and proper comparison between different measurements. Maximum of the signal measured (resp. minimum) corresponds to a negative peak of second derivative (resp. positive), used to numerically perform peak detection afterwards. The green dashed lines are guides for the eye showing resonances found by data analysis with automatic peak detection routine on the smoothed second derivative of the normalized signal. At frequency of 112 GHz, magneto-photoconductivity shows five lines, two at negative magnetic fields ($B_{-1} = -3.422$ T $\pm 8$ mT and $B_{-3} = -4.320$ T $\pm 5$ mT) and three at positive magnetic fields ($B_{+1} = 3.507$ T $\pm 11$ mT, $B_{+2} = 3.975$ T $\pm 18$ mT and $B_{+3} = -4.215$ T $\pm 22$ mT). Resonance fields are obtained from the mean value of the peak position detected for each back gate voltage curve, while the uncertainty is estimated from the standard deviation. Note that $B_{-1}$ and $B_{-3}$ lines are almost symmetric to $B_{+1}$ and $B_{+3}$ with less than 100 mT deviation, to be compared to resonance width roughly estimated at 200 mT. Note also that a small deep in signal is visible around $-4.0$ T, indicated in grey dashed line for clarity.

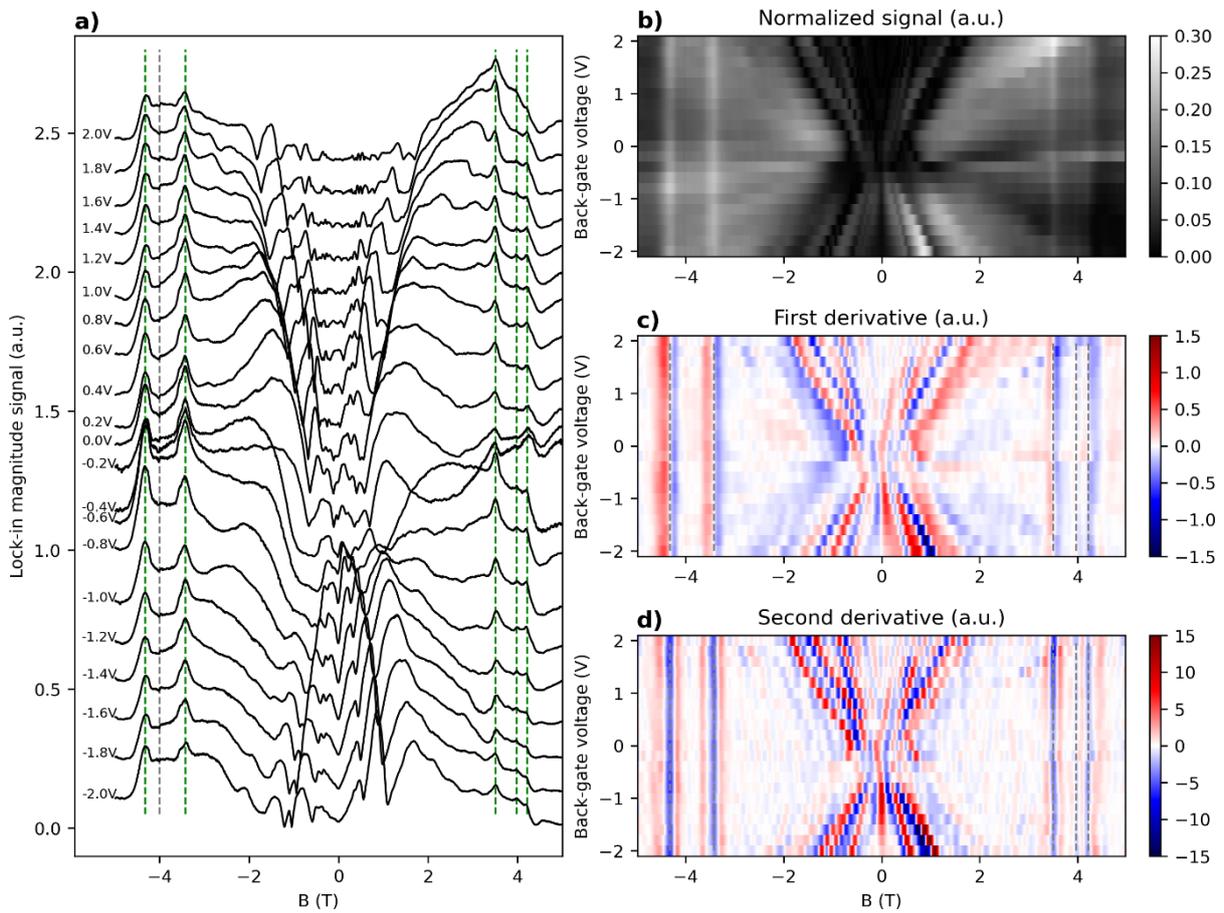

Figure 1 − a) Measured photoconductivity signal in sample B at a given incident frequency and temperature ($f$ = 112 GHz, $T$ = 4 K) as a function of magnetic field, for different back gate voltages. The X-shaped feature at low field indicates the presence of Shubnikov-de Haas-like oscillations shifting towards stronger magnetic fields as the gate voltage increases. At higher magnetic field, several transitions are well seen as vertical lines, meaning that there are not affected by the back gate voltage. Green dashed lines correspond to the mean value of numerical peak detection routine, while grey dashed line at –4 T is just indicative added manually. b) Absolute value of normalized signal represented as the colour map plot. Signal is normalized to the integral over B. c) First derivative with respect to magnetic field. Each absorption line results in a dispersive shape curve with polarity change at the maximum. d) Second derivative with respect to magnetic field. Maximum of signal (resp. minimum) corresponds to a negative peak of second derivative (resp. positive), used to numerically perform peak detection.

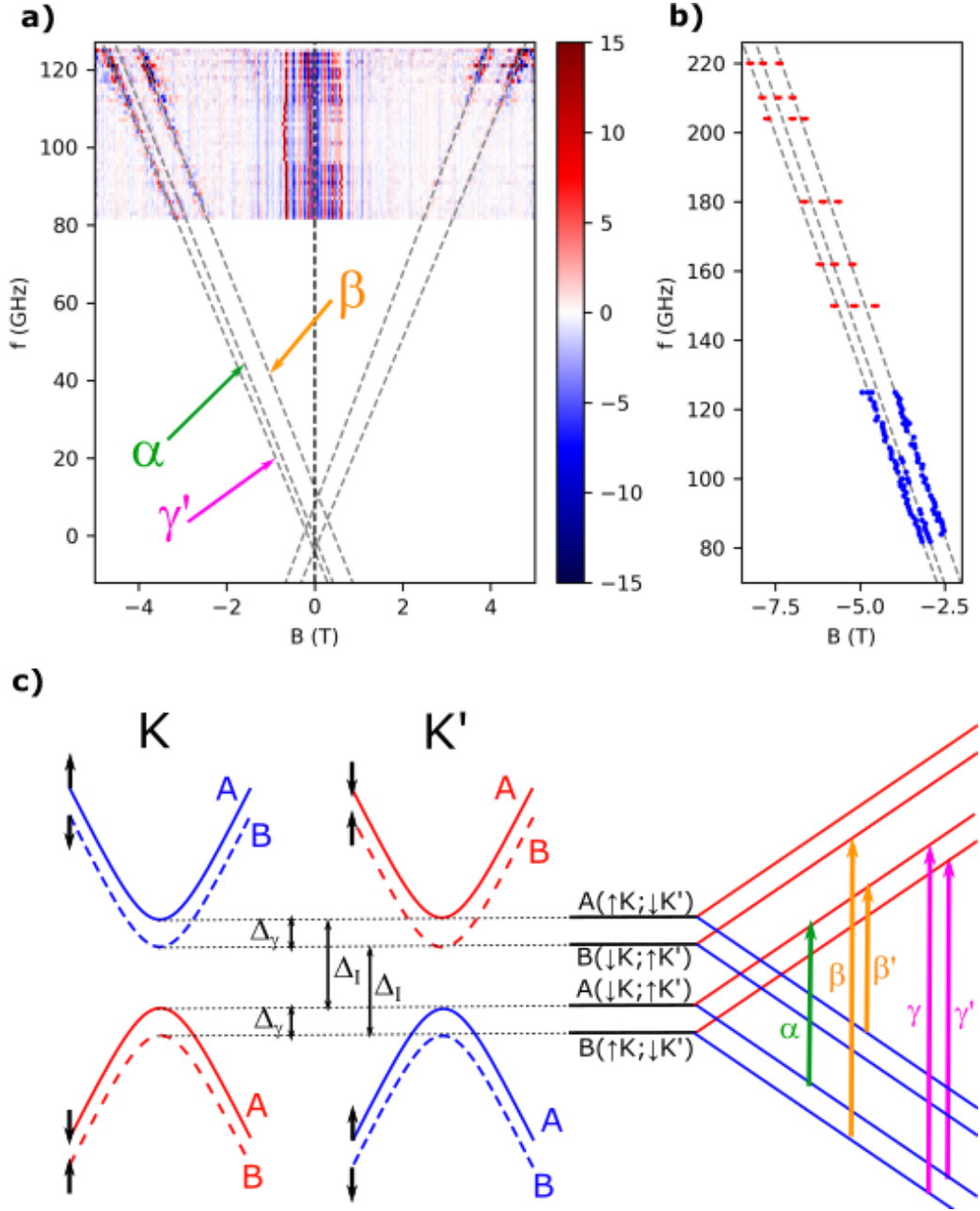

Figure 2 – a) Second derivative of the signal measured in sample B as a function of the magnetic field at 2.6 K in the frequency range 82-125 GHz. b) Frequency of the different resonances as a function of magnetic field in the range 82-220 GHz measured at 4 K. The dashed lines are linear fits obtain from measurement in the 82-125GHz range allowing for the extrapolation of the experimental results to B = 0. The highest dashed line (β resonance) follows the linear evolution with a slope of -29.3 GHz/T ± 0.4 GHz/T and with an intercept at B = 0 of +11±1.0 GHz (45.7±8 μeV). The two other dashed lines corresponds to the α (resp. γ') resonance and have almost the same slope of -28.2 GHz/T ± 0.4 GHz/T (resp. -27.4 GHz/T ± 0.3 GHz/T) but their intercepts extrapolate to -1.1±1.0 GHz (-4.5±4.1 μeV) and –7.0±0.9 GHz (–28.8±3.7 μeV) at B = 0, respectively. These values are consistent with ISO and sub-

lattice ZFSs. c) Gapped dispersion relation of graphene at *K* and *K'* points of the Brillouin zone for the atoms of sites A and B (left side). The black horizontal lines correspond to the four levels whose degeneracy is lifted by the ISOs and sub-lattice potentials. When a magnetic field is applied (on the right side), the levels are splitted due to the Zeeman effect. The allowed spin-flip transitions are marked by coloured arrows.

This decrease of conductivity is too small (in width and intensity) to be retrieved properly with a numerical peak detection routine, affected by long term drifts of the signal. It could be however interpreted as a sixth line, $B_{-2} \approx -4$ T, in coherence with the line $B_{+2}$.

We then studied the evolution of these different resonances as a function of the excitation frequency. The frequency of these spin resonances scales linearly with the magnetic field from 45 GHz up to 220 GHz. Fig. 2 a) shows an example of the obtained results in the range 82-125 GHz on sample B (bilayer graphene). For each working frequency, the signal is processed as follow. First, the magnitude of the lock-in output measured is smoothed with a standard Savitzky-Golay filter, from *scipy.signal* Python library, with a window width typically of 50 mT. Then, this signal is numerically derived in regard to the magnetic field *B* using *numpy.diff* python method, followed once again by a smoothing from a Savitzky-Golay filter with a window width typically of 50 mT. This first derivative is then derived again in regard to the magnetic field *B* and smoothed with a Savitzky-Golay filter, and denoted afterwards the second derivative $\frac{\partial^2 S}{\partial B^2}$. A peak of photoconductivity has a second derivative form itself by a narrower peak but at the same position, with two side peak four times lower in amplitude. These peak positions are automatically extracted from those data using *find_peak* routine of *scipy.signal* library in Python, with typically prominence parameter of 0.5, and peak distance of 100 mT. Local maxima appear then in red while local minima are plotted in blue. For magnetic fields lower than 1 T in magnitude, those peaks measured in the signal are independent of the working frequency and they are interpreted as Shubnikov-de Haas-like oscillations with no dependence with the incident wave frequency.

For larger magnetic field, several lines with a linear dependence with the frequency and magnetic field are clearly visible (grey dashed lines in Figure 2 (a)). For each frequency, the position in magnetic field of each peak is automatically extracted by the peak detection routine. These three spin resonances can be extrapolated to zero magnetic field either by adjusting the results of data processed automatically by the peak detection routine (Fig. 2a), or by placing the position of the resonances observed by eye as a function of the frequency within the whole frequency range used in this study (Fig. 2b). In order to interpret quantitatively these measurements, we used the same theoretical description than reference [6]. where the Hamiltonian of the system at $B = 0\,T$ is based on a minimal model in the bispinor basis spanned by spin and sublattice spin $\{\uparrow, \downarrow\} \otimes \{A, B\}$,

$$H = \hbar v_F I \otimes \ \tau \sigma_x k_x + \sigma_y k_y \ \ + \frac{1}{2}\Delta_I \tau s_z \otimes \sigma_z + \frac{1}{2}\Delta_\gamma s_z \otimes I,$$

where $\Delta_I$ is the ISO gap, $\Delta_\gamma$ the sublattice splitting, $\tau$ the valley index, $k_x, k_y$ the small vector near the Dirac point, and $I$ the identity operator. In that model, $\Delta_I$ and $\Delta_\gamma$ are assumed to depends only on the temperature.

For sake of clarity, Fig. 2 c) shows a diagram of the splitted bands of graphene at $K$ and $K'$ points of the Brillouin zone at zero magnetic field for the atoms of sites A and B. The usual Dirac cones of opposite chirality (represented in red and blue) have opposite gaps in $K$ and $K'$. In the case of symmetric sub-lattices, each of its bands is doubly degenerated ($\Delta_I$). But when considering a sub-lattice symmetry breaking driven by the interaction of graphene with the hBN substrate, a second rise of spin degeneracy takes place between the two sub-lattices A and B (see dashed lines) ($\Delta_\gamma$). In the presence of magnetic field (right part of panel c), each of these four bands is Zeeman splitted. Several optical spin-flip transitions are therefore allowed. The $\alpha$ resonance, corresponding to the central feature in Fig. 2a) and b), is the ordinary Zeeman splitting $E_Z = g\mu_B B$. The upper resonance $\beta$, also reported in Fig.2a) and b), reveals a ZFS which is a direct measurement of the ISO coupling gap $\Delta_I$. The lowest resonance $\gamma'$ corresponds to the ZFS attributed to the sub-lattice splitting $\Delta_\gamma$ due to the coupling of the graphene sheet with the hexagonal boron nitride (hBN) encapsulation [6].

In Figure 2a), dashed grey lines correspond to a linear fit of the peak position for each line. From each linear fit, one extracts the Landé factor $g$ from the slope, while the intercept at zero magnetic field provides the value of the zero-field splitting of the corresponding spin-flip transition. For the $\alpha$ resonance, the slope is measured at -28.2 GHz/T ± 0.4 GHz/T and an intercept at B=0T of -1.1±1.0 GHz (-4.5±4.1 μeV). For the $\beta$ resonance, the slope is measured at -29.3 GHz/T ± 0.4 GHz/T and an intercept at B=0T of 11.0±1.0 GHz ($\Delta_I$=45.7±4.1 μeV). For the $\gamma'$ resonance, the slope is measured at -27.4 GHz/T ± 0.3 GHz/T and an intercept at B=0T of -7.0±0.9 GHz ($\Delta_\gamma$=-28.8±3.7 μeV). These results are summed up in Table $I$ and in Supplementary Material.

| Resonance | Slope (GHz/T) | Frequency at B=0 (GHz) |
|---|---|---|
| $\gamma'$ | $-27.4 \pm 0.3$ | $-7.0 \pm 0.9$ |
| $\alpha$ | $-28.2 \pm 0.4$ | $-1.1 \pm 1.0$ |
| $\beta$ | $-29.3 \pm 0.4$ | $11.0 \pm 1.0$ |

Table 1 - slope and frequency at B=0T extracted from data of Fig. 2a).

For single layer graphene (sample A), the Landé factor, calculated from the slope, was measured to be $g = 2.09\pm0.02$ (more information can be found in supplementary materials), and $g = 2.02\pm0.02$ for bilayer graphene sample. In Fig. 2b), due to low signal obtained at high frequencies, positions of the resonances have been manually noted (the red filled circle in Fig.2b)). However, the results are consistent with these linear fit in grey dashed lines (linear fit are realized only on blue filled circle, in the 82 – 125 GHz frequency range).

The extrapolation of the central feature ($B_{-2}$ or $B_{+2}$) intersects with the axis at its origin, when the other two lines extrapolate to finite positive and negative energy values (Fig. 2a). Similar results were also obtained with approximately the same resonant energies in sample A, on which we observed such resonances down to 45 GHz (see Supplementary Materials), and similarly in monolayer and trilayer graphene samples studied elsewhere [5,6,7]. In these works, the central feature represents the ordinary Zeeman splitting and the upper and lower lines were identified as spin-flip transitions between the splitted bands. Values of Landé g-factor, $\Delta_\gamma$ and $\Delta_I$ splitting are summed up and compared to reference [6] in Table 2.

| | Landé g-factor | | | | $\Delta_\gamma$ (µeV) | $\Delta_I$ (µeV) | $hf_{0,\alpha}$ (µeV) |
|---|---|---|---|---|---|---|---|
| | $\alpha$ | $\beta$ | $\gamma$' | Mean value | | | |
| Sample A | 2.01 ± 0.03 | 2.09 ± 0.03 | 1.96 ± 0.02 | 2.02 ± 0.02 | 25.3 ± 6.1 | 51.1 ± 6.5 | $-1.6 \pm 5.7$ |
| Sample B | 2.00 ± 0.03 | 2.025 ± 0.04 | 1.98 ± 0.03 | 2.00 ± 0.02 | 28.8 ± 3.7 | 45.7 ± 4.1 | $-4.5 \pm 4.1$ |
| Ref. [6] | 1.90 ± 0.04 | - | - | 1.90 ± 0.04 | 18.69 ± 0.62 | 45.24 ± 0.79 | - |

Table 2 – Landé g-factor, $\Delta_\gamma$ and $\Delta_I$ extracted from measurements on sample A (single layer graphene) and sample B (bilayer graphene). The value of the a resonance frequency $f_{0,\alpha}$ intercept at B=0T field, noted $f_{0,\alpha}$, is also indicated in energy unit ($hf_{0,\alpha}$). Values from Ref. [6] are indicated for comparison.

From level scheme of Fig. 2c), one would expect five resonance frequencies for a given magnetic field $\alpha$, $\beta$ and $\beta$', $\gamma$ and $\gamma$'. For each of these resonances, the Zeeman splitting provides a linear dependence in magnetic field, so that the frequency at zero field (B=0T), noted $f_0$, might be estimated from the intercept of the linear fit of experimental data. The zero field resonance frequency is expected to be null for the $\alpha$ resonance ($f_{0,\alpha} = 0$ GHz), while for the four other ones is directly related to $\Delta_I$ for $\beta$ and $\beta$'resonances ($f_{0,\beta} = \Delta_I/h$ and $f_{0,\beta'} = -\Delta_I/h$), and to $\Delta_\gamma$ for $\gamma$ and $\gamma$' resonances ($f_{0,\gamma} = \Delta_\gamma/h$ and $f_{0,\gamma'} = -\Delta_\gamma/h$). However, one observes only three resonances ($\alpha$, $\beta$ and $\gamma$') over the five expected, attributed to different strength so that resonances $\beta$' and $\gamma$ are not visible within the signal-to-noise ratio of the experimental setup used. Such difference in resonance strength has been reported in reference [6] where only $\alpha$, $\beta$ and $\gamma$' resonance have been clearly measured from several frequencies while $\gamma$ resonance has been observed for few frequencies with very low signal-to-noise. Even though the value of $\Delta_\gamma$ is not exactly the same as found in previous work ($\sim$ 30 µeV here, instead of 20 µeV in Ref. [6]), we believe that this gap is due to the same physical phenomenon. Indeed, as it is not an intrinsic value of graphene, there is no reason why these values should be exactly the same in samples with different geometries. However, the value of $\Delta_\gamma$ is very comparable in our samples A and B, even if they were processed in different ways (See SM for details). Additionally, we also compare the spin relaxation times $\tau_s = \hbar \left( 2h \frac{\Delta f}{\Delta B} \delta B \right)^{-1}$ in our samples to those obtained in previous work [6]. For this, we use the β transition from figure 2, for which $\Delta f/\Delta B$ is the slope extracted from the linear fit in figure 2 and $\delta B$ the resonance peak width. We thus obtain a spin relaxation time in the range of ~17 – 25 ps within the temperature range 2.6 –8 K, which is consistent with previous results [6].

Finally, we address the question of the behaviour of these different electron spin resonances as a function of temperature. The experimental study was carried out between 2.6 K and 12 K. Figure 3a) shows the photoconductivity signal for sample A, measured at different temperatures as a function of magnetic field, with an incident frequency of 60 GHz. The probe frequency has been chosen such that resonances are observable up to at least 10K. For each temperature, several relevant peaks appear in the signal in the middle of other structures induced by different conductance oscillations and other measurement noises. Relevant peaks are clearly visible in typical frequency and magnetic field maps provided in Fig. 2, where their position dependence in frequency helps to clearly identify them, while the noise-related peaks are randomly distributed with frequency. Then, we used the linear fit parameters extracted from data, and their confidence interval, of Fig. 2 to predict the value of magnetic field, at which one expects to have a spin resonance for the working frequency used (60 GHz). Doing that, one may estimate the position of the three expected lines at 2.6 K and the corresponding uncertainty at −2.44 T±0.05 T, −2.16 T±0.05 T and −1.67 T±0.05 T.

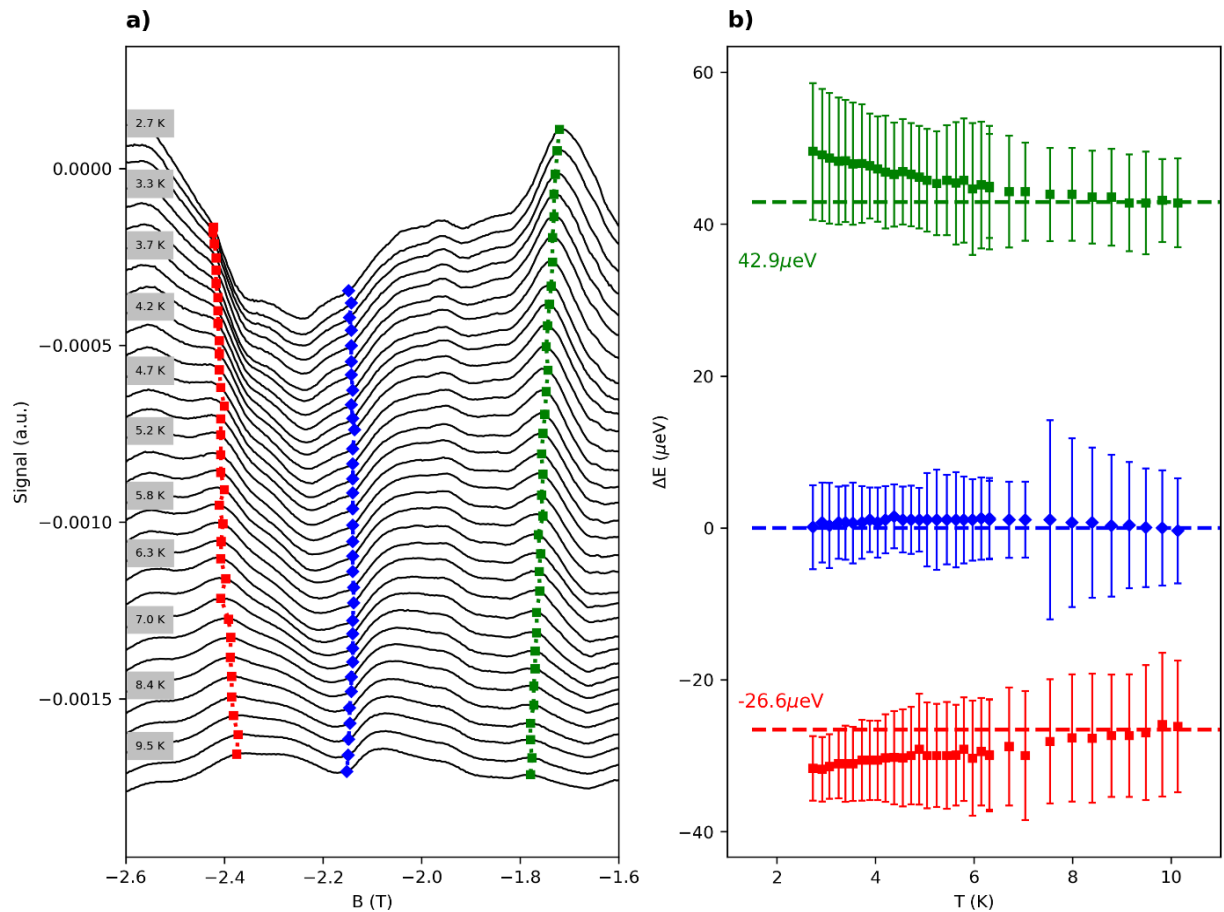

Figure 3 - Evolution of the resonance in sample A versus temperature for a source frequency of 60 GHz. a) Photoresponse is plotted versus magnetic field at different temperatures. Red and green square represent the evolution with the temperature for local maximum of signal while blue diamond

represents local minima of signal. From previous measurements of photoconductivity as a function of magnetic field and frequency (see **Ошибка! Источник ссылки не найден.**), one may estimates the position of the three expected lines and corresponding uncertainty : $-2.44\text{T} \pm 0.05\text{T}$, $-2.16\text{T} \pm 0.05\text{T}$ and $-1.67\text{T} \pm 0.05\text{T}$. b) Energy of the intercept at zero origin versus the temperature for the three observed transitions. Same color as a) is used to represent the transition. Dash line represents the converge values of energy at high temperature, corresponding to 42.9 and -26.6 µeV respectively. Error bars are estimated on the second derivative signal, using the *peak_widths* routine from Python *scipy.signal* library, with a relative height parameter (*rel_height*) of 0.5.

The uncertainty is sufficiently low to identify unambiguously each line of interest in Fig. 2. These peaks are marked with different symbols on the raw data in Fig.3a). The corresponding spin resonance energy is calculated from the line positions in magnetic field, with the mean value of the α line taken as the origin of the energy. The Zeeman shift is therefore renormalized so that the β and γ lines correspond directly to the ZFS. so that β and γ lines are compensated from Zeeman shift to the zero-field splitting. We assume a linear Zeeman shift with a similar Landé factor for each line, as observed in previous results (see Fig.2 and SM). A clear effect of temperature is observed on the position of resonances, which shift away from each other as the temperature decreases. This temperature dependence of the position of the resonances was observed in our two samples at several frequencies (cf. Fig. SM 2, see, also, references [35,36]). In both samples, the $\alpha$ resonance was always represented the weakest line, which position was consistently independent of temperature.

By extrapolating to zero magnetic field the positions of the resonances measured at 60 GHz, we find the values of the different ZFS. Their evolution in temperature is represented in Figure 3 b). It can be seen that the ISO ZFS increases from 45 µeV to 55 µeV between 12 K and 2.6 K, while the ZFS due to the sub-lattice potential increases from 30 µeV to 40 µeV in the same temperature range.

## Discussion

To the best of our knowledge, this modification of the ZFS driven by temperature has never been reported in graphene. A similar effect was however observed at higher temperatures in copper II dimer [37], where it was supposedly induced by the dynamic Jahn–Teller effect, and in hBN nanopowders [38] in which it was attributed to the thermal expansion. Additionally, a similar a similar temperature dependence was also observed at lower temperatures in several magnetic materials such as $CrCl_3$ [39] or $BaAg_2Cu[VO_4]_2$, resulting from ferromagnetic and antiferromagnetic states. Therefore, a first way of interpreting our experimental observation might be to consider the presence of a magnetic order at low temperature [40] likely to modify the influence of the external magnetic field on the spin resonances. However, this interpretation is unlikely for several reasons. Firstly, the β and γ' spin resonances evolve towards opposite magnetic fields, while the α resonance remains independent of temperature. Secondly, the temperature evolution of the ESR intensity signal, $\chi_{ESR}$, which is directly proportional to the spin susceptibility [41], shows no trace of magnetic order. The signal seems indeed to obey Curie's law with a Curie-Weiss constant tending to zero (Fig. SM 3 in Supplementary

Information). It is conceivable that the staggered sub-lattice potential can be modified by a temperature-driven strain of the graphene layer, as soon as the sub-lattice symmetry is broken by the interaction of the graphene layer with its hBN substrate. For instance, it was evidenced in [42] that the strain caused by thermal expansion coefficient mismatch between graphene and substrate cannot be neglected when compared with suspended graphene. In [43] authors found that when cooling graphene from 300 K to 10 K, the influence of strain on the monolayer and top and bottom layer of the bilayer graphene is large and shown a pronounced temperature-dependent variation. But the idea seems less natural with regard to the ISO which represents an intrinsic parameter of graphene sheets. One would indeed tend to say at first glance that the amplitude of the ISO should not depend on the temperature since it is a purely relativistic phenomenon due to the motion of electrons around carbon atoms. However, it was also shown in [44] that geometric curvature of the graphene sheet should affect the spin–orbit coupling. Later, B. Gong *et al.* [45] have shown using tight-binding approach that uniaxial strain can be used as a reversible and controllable way to tune the ISO coupling in graphene. In the case of applied uniaxial strain, not only the change in atomic distances has to be taken into account but also the lattice deformation, which affects the orbitals reorientation. The dependence of ISO splitting on the type of the strain is theoretically predicted by means of first-principles calculations in [46]. Additionally, by using a tight-binding model, it has been found that the strain should make it possible to control the strength of Rashba and intrinsic spin-orbit coupling [47]. And very recently the thermal expansion coefficient of bilayer and trilayer graphene was finally measured [48]. In this work, the authors claim that the metal deposit may cause local strain in 2D materials around metal elements such as contacts and top gates, even though edge contacts were found in there to have no significant impact on the resulting strain. However, A. De Sanctis et al. have shown experimentally in a twist-angle Graphene/hBN device [49] that top-contacts induce a complex strain pattern in the graphene layer. Indeed, as the thermal expansion coefficients of gold and graphene are opposite, this leads to complex contraction behavior upon cooling depending on the overall geometry and design of the graphene-based device. In our case, the device are dual grating top-gate structures composed of long parallel metal fingers dedicated to the coupling of the incident THz radiation to the 2D electron system (See Fig. M1 in Methods). The thermal expansion difference between the graphene flake and the metallic grating should certainly be taken into account. Therefore, the observed variation of the ZFS with temperature may tentatively be attributed to the strain thermally induced from the metallic grating to the monolayer and bilayer graphene on hBN in both samples A and B. In such a case, the strain would modify on the one hand the distance between the carbon atoms, which would act on the staggered sublattice potential and the position of the $\gamma$ line. On the other hand, this would distort the lattice, which would induce a reorientation of the orbitals, modifying the strength of the ISO and displacing the $\alpha$ line. However, it is essential to point out that the effects of thermal expansion or contraction, previously observed in graphene and reported in the literature [50, 51] are systematically observed at higher temperatures than those used in our experiments. At cryogenic temperatures, we can indeed assume that everything should already be strained when cooling from 300 K to 100 K, and that we should no longer expect an evolution below these temperatures. An alternative explanation could therefore be related with the effect of the electron-phonon interactions, often involved in temperature dependencies of the electronic structure in semiconductors and insulators. For instance, H. Ochoa and co-workers analysed the possible couplings between spins and flexural, out-of-plane, vibrations in graphene and found that the coupling with the phonons, should renormalize notably the Kane-Mele mass [16]. It was shown soon after, that supercollision scattering processes, facilitated by ripples, or flexural modes, are the dominant mechanism of electron-phonon energy transfer in suspended graphene [52]. Moreover, Kurzmann et al. recently studied graphene quantum dots and argued that the spin-orbit coupling of nominally flat graphene is enhanced by the out-of-plane zero-point vibrations of graphene [53]. On another side, Norambuena et al. modelled the effect of phonons and temperature on the ESR spectrum in molecular systems in the presence of a Jahn-Teller effect. They calculated the thermal dependence of Ham reduction factors [54] and showed its influence on the energy of a ZFS. Although their model is not specifically adapted to the case of graphene, it can be noted on the one hand that their theoretical results present qualitative similarities with the behaviour observed in our samples, and on the other hand the vacancy of a carbon atom in the crystal lattice of graphene can indeed be reconstructed by means of a Jahn-Teller distortion [55,56]. However, our results do not seem to indicate signs of quenching of the spin-orbit coupling predicted by their model, but rather a saturation of its value beyond 8 K.

Additional experimental and theoretical studies are needed to elucidate the origin of the temperature evolution of ZFS. An experimental way to investigate these hypotheses in more detail would be for example to study the influence of the hydrostatic pressure on the ISO in a similar experimental configuration.

## Conclusion

In conclusion, by means of a sub-Terahertz photoconductivity-based electron spin resonance technique, we have investigated the influence of temperature on different spin-splittings in monolayer and bilayer graphene. We have observed three main electron spin resonances systematically in our two Ratchet THz detectors over a frequency range from 45 GHz to 220 GHz. By extrapolating these resonance frequencies at zero field, we extracted two ZFSs, attributed to ISO and sublattice potential, whose values are comparable to those in the literature. By varying the temperature of these two samples, we found that these two ZFS increase by about 25 % when the temperature drops from 12 K to 2.6 K. We attempt to interpret these results by considering successively the possible presence of a magnetic order, the strain effect induced by the difference in thermal expansion between the graphene and the top gate contacts, and finally the electron-phonon interactions. None of these hypotheses is completely satisfactory, therefore the origin of our experimental observation requires further theoretical and experimental investigations. Beyond this particular behaviour, we also validate the fact that graphene-based photoconductive THz detectors allow efficient measurement of electronic spin resonances at high frequencies and magnetic fields. Indeed, even if the signal was higher in the ratchet detectors, the spin resonances were clearly observed in both types of THz sensors.

## Samples and methods

**Samples**: Our three samples are made of monolayer (sample A and C) and bilayer (sample B) graphene. The graphene layers in samples A and B are encapsulated with relatively thin hBN flakes on a SiO2/Si substrate (300 nm SiO$_2$) by micromechanical exfoliation of bulk materials using a standard scotch tape method. In addition, the devices also include a double metallic interdigitated asymmetric top gate (DGG) structures. These multiple gate periodic structures have been widely studied as broadband THz sensors in graphene based systems, where the photocurrent generated in the graphene channel is due to the well-known ratchet effect [57, 58]. The grating is formed by independent wide (TG$_2$) and narrow (TG$_1$) strips which allows to apply different top gate potentials. A schematic view of the fabricated device is shown in Figure M.1 (a). The optical image for the bilayer graphene structure is shown in Fig. M.1 (b). More detailed information on the sample fabrication can be found in Refs. [57] and [59]. Sample C (see schematics in Figure M.1 (c) is a THz detector based on a monolayer graphene p-n junction with log-periodic antenna used to couple the incident THz light. Note that the gate voltage corresponding to the CNP, UCNP, was varying slightly between different sample cool downs due to different charge trapping in the gate insulator [60,61].

The graphene detectors with thermoelectric readout (involving p-n junctions) were fabricated by using chemically-vapor-deposited (CVD) graphene on a 2" large copper foil either 25 or 60 μm thick in the commercial cold-wall system (AIXTRON Black Magic II). The charge-carrier mobility of such a graphene transferred to ordinary office lamination foil (EVA/PET) was surprisingly high, reaching 9000 cm$^2$/(Vs) at room temperature [62]. The p-n junctions were fabricated in the center of the graphene channel by chemical doping (see Fig. M.1c) and were assumed to be also exist near the metal electrodes through the proximity doping.

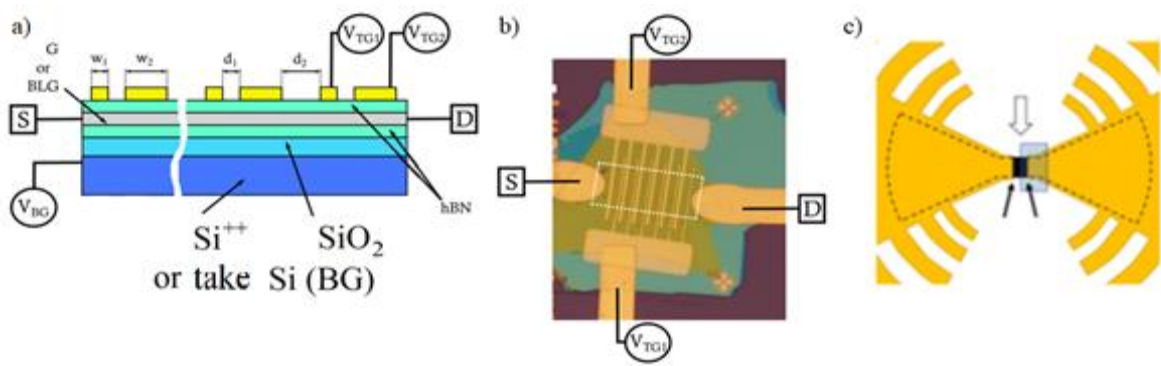

Figure M.1 - Device description. **a)** The devices A and B are characterized by a periodic unit cell, L, of $w_1 = 0.5$ μm, $w_2 = 1$ μm, $d_1 = 0.5$ μm, $d_2 = 2$ μm for the Sample A and $w_1 = 0.75$ μm, $w_2 = 1.5$ μm, $d_1 = 0.5$ μm, $d_2 = 1$ μm for the Sample B. **b)** Picture of ratchet THz sample B where the encapsulated bilayer graphene has been highlighted by the dashed line **c)** The model geometry of graphene detector with a p-n junction in the center marked by the empty arrow. Graphene is outlined by the dashed line. The rectangle shows a layer of photoresist covering one half of the graphene channel. The solid arrows mark two latent p-n junctions in the vicinity of the electrodes (log-periodic antenna in this case).

**Measurements**: The samples are placed inside a 6T horizontal cryogen free magnet system with optical access (See Figure M.2). The voltage generator (Keithley 2600B) allows to control the voltage applied on the back-gate with a voltage on the top-gate fixed at 0V for the ratchet samples A and B. Sub-THz source generated by Shottky diode of multiplied frequency are used to obtain frequencies in the ranges from 45 GHz to 75 GHz (optical power about 150 mW), and 82 GHz to 125 GHz (optical power about 10 mW). The sub-THz beam is focalized on the sample, through three Z-cut quartz windows, with golden parabolic mirrors. A magnetic field up to 5.5 T is oriented perpendicular to the surface of the sample and parallel to the wave-vector of the incident radiation. The signal is detected as a voltage drop at the edges of the detector, which is then amplified and measured via a standard lock-in technique (using an Amatek Signal recovery 6270). Similar setup on a 16T vertical magnet system is used to obtain results at higher frequencies with a sub-THz source from 150 GHz up to 220 GHz (power source of 5 mW).

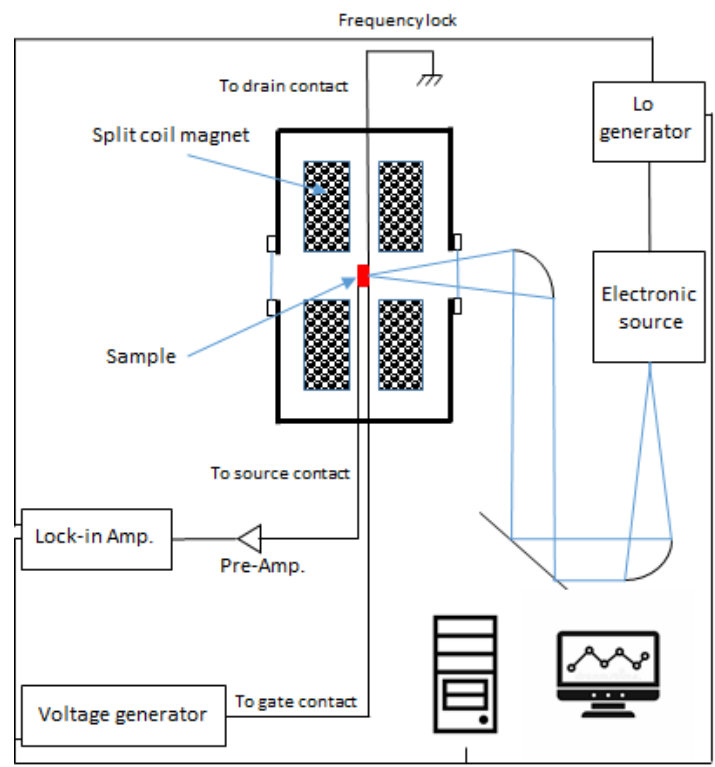

Figure M.2: Description of photoconductivity measurement. The sample is placed inside a 6T horizontal cryofree magnet system. Sub-THz sources are used to illuminate the sample and excite

resonantly the spin-flip transitions. The photoconductivity signal is measured via a standard lock-in technique.

**Detection principle:** In the two encapsulated graphene samples with the DGG structure, the THz radiation results in the ratchet current, which is caused by the combined action of a spatially periodic asymmetric in-plane potential and the spatially modulated light due to the near-field effects of the radiation diffraction [57,58,63]. The amplitude and the sign of the photocurrent induced in graphene is defined by the lateral asymmetry parameter given by

$$\Xi = \overline{|\boldsymbol{E}(x)|^2 \frac{dV(x)}{dx}}$$

Here the overline stands for the average over the ratchet period, $dV(x)/dx$ is the derivative of the coordinate-dependent electrostatic potential $V(x)$, and $\boldsymbol{E}\ x$ the distribution of the radiation electric field being coordinate dependent due to the near field of diffraction. In this study both top gates are kept at zero bias, however, the asymmetry is created by the built-in potential due to the metal stripes of different widths (TG1 and TG2) deposited on top of the encapsulated BLG. Note that in sample B the zero-magnetic field ratchet effect was studied in detail in Ref. [57].

With a single p-n junction formed in the graphene channel by a split gate or chemical doping, the response (dc) signal $V_s$ appears because of a non-uniform Joule heating and thermoelectric effects, $V_s \sim S\ x\ \nabla T(x)$. The Peltier effect will either help the Joule heating or reduce it, depending on the direction of current through the p-n junction. By averaging the instant signal voltage over one period of the THz radiation, one finds the mean value of the signal, which is measured in the experiment. In the case of no intentionally made p-n junction in the graphene channel, two latent p-n junctions can still exist because of extra doping through the proximity to metal electrodes. These p-n junctions normally do not contribute to the output signal because they are connected back-to-back and their individual contributions to the signal compensate each other. However, in the presence of a small dc current, one of the junctions will be heated more than another because of the Peltier effect. This will break the symmetry of the device and result in a non-compensated signal [64].

## Acknowledgements

This work was supported by the Terahertz Occitane Platform, by CNRS through IRP "TeraMIR", by the French Agence Nationale pour la Recherche (Colector-ANR-19-CE30-0032, Stem2D-ANR-19-CE24-0015 and Equipex+ Hybat-ANR-21-ESRE-0026 projects), by the European Union through the Flag-Era JTC 2019 - (project DeMeGRaS, VR2019-00404, DFG No. GA 501/16-1, and ANR-19-GRF1-0006) and the Marie-Curie grant agreement No 765426, from Horizon 2020 research and innovation programme. FT, SDG, WK, JADN and IY thank the support from the IRAP program of the

Foundation for Polish Science (grant MAB/2018/9, project CENTERA). EM and SDG acknowledge the Volkswagen Stiftung Program 97738. YMM acknowledges the support by the Spanish Ministry of Science, Innovation, and Universities and FEDER under the Research Grants numbers RTI2018-097180-B-100 and the FEDER/Junta de Castilla y León Research Grant number SA121P20. JADN thanks the support from the Universidad de Salamanca for the María Zambrano postdoctoral grant funded by the Next Generation EU Funding for the Requalification of the Spanish University System 2021–23, Spanish Ministry of Universities.

## Author contributions

The experiment was proposed by FT. The samples were fabricated by JADN, IY, EM, and KI. THz photoconductivity experiments were carried out by CB and CC. Characterization measurements were conducted and analysed by BJ, SG, KD, SDG and SR. KM and CB handled the data and prepared the figures. FT, and KM wrote the manuscript and CB, JADN, SSK, AY, YMM, WK, SSK and SDG corrected it. All co-authors discussed the experimental data and interpretation of the results.

## Competing interests

The authors declare no competing financial interests.

## Data availability

Data are available upon reasonable request to the corresponding author. Dataset will also be uploaded on the Recherche Data Gouv repository (https://recherche.data.gouv.fr) once the manuscript is accepted and published.

# Supplementary Information

# Temperature Dependent Zero-Field Splittings in Graphene

C. Bray[1,‡], K. Maussang[2,‡], C. Consejo[1], J. A. Delgado-Notario[3,4], S. Krishtopenko[1], I. Yahniuk[4,5], S. Gebert[1,2], S. Ruffenach[1], K. Dinar[1], E. Moench[5], J. Eroms[5], K. Indykiewicz[6], B. Jouault[1], J. Torres[2], Y. M. Meziani[3], W. Knap[4], A. Yurgens[6], S. D. Ganichev[4,5], F. Teppe[1*]

[1] L2C (UMR 5221), Université de Montpellier, CNRS, Montpellier, France

[2] IES, (UMR 5214), Université de Montpellier, CNRS, Montpellier, France

[3] Nanotechnology Group, USAL-Nanolab, Universidad de Salamanca, Salamanca 37008, Spain

[4] CENTERA Laboratories, Institute of High Pressure Physics, Polish Academy of Sciences, 29/37 Sokołowska Str, Warsaw, Poland

[5] Terahertz Center, University of Regensburg, 93040, Regensburg, Germany

[6] Chalmers University of Technology, SE-412 96 Göteborg, Sweden

* frederic.teppe@umontpellier.fr

[‡] These authors have contributed equally to this work and share first authorship

## Data processing

In order to increase visibility of the absorption lines in the signal, the following data treatment has been applied to raw photoconductivity data. The signal, denoted hereafter $S$ is the magnitude $R$ measured by the lock-in. Firstly, this is signal is smoothed with a standard Savitzky-Golay filtre, from scipy.signal Python library, with a window length of typically 50m T to 100m T and a 3rd order polynomial interpolation. This smoothed signal is renormalized to unity with a norm 2 metric.

The derivative signal is then calculated from the finite difference of the normalized smoothed signal, denoted $D_1$, where the derivation is taken regarding the magnetic field $B$. This treatment allows removal of parasitic DC offsets and long-term drifts of signal during the experimental mapping.

$$D_1 = \frac{\partial S}{\partial B}.$$

Assuming a Lorentzian absorption line, centered at $B_0(f)$ where $f$ is the incident wave frequency and with a FWHM of $2\Gamma$, the derivative signal will be a dispersive curve as follows

$$S(B, f) = \frac{a}{1 + \left(\frac{B - B_0(f)}{\Gamma}\right)^2},$$

$$D_1 = \frac{\partial S}{\partial B} = -\frac{2a}{\Gamma^2} \frac{\left(B - B_0(f)\right)}{\left[1 + \left(\frac{B - B_0(f)}{\Gamma}\right)^2\right]^2},$$

This derivative is then smoothed with a Savitzky-Golay filter with a window length of typically 50mT to 100mT and a 3rd order polynomial interpolation. Then, the second derivative is numerically computed so that a maximum of signal corresponds to a maximum of second derivative, denoted $D_2$, while quadratic drifts of signals are reduced

$$D_2 = \frac{\partial^2 S}{\partial B^2} = -\frac{2a}{\Gamma^2} \frac{1 - 3\left(\frac{B - B_0(f)}{\Gamma}\right)^2}{\left[1 + \left(\frac{B - B_0(f)}{\Gamma}\right)^2\right]^3},$$

This $D_2$ signal is then extremal for $B = B_0(f)$ with a width based on first annulation of $\frac{2\Gamma}{3}$.

Let's assume a signal of the following expression

$$S(B,f) = \frac{a}{1 + \left(\frac{B - B_0(f)}{\Gamma}\right)^2} + h(B,f),$$

where $h(B,f)$ is a function that models technical drifts and baseline of the experimental setup. Peak localization on $S(B,f)$ is already affected by the B field dependence of $h(B,f)$ since the first derivative is expressed as follow

$$D_1 = \frac{\partial S}{\partial B} = -\frac{2a}{\Gamma^2} \frac{\left(B - B_0(f)\right)}{\left[1 + \left(\frac{B - B_0(f)}{\Gamma}\right)^2\right]^2} + \frac{\partial h}{\partial B}.$$

Peak localization of the resonance on the initial signal is given by the cancellation of the first derivative, providing $B_{measured}(f)$ and then is affected by the first derivative of $h$. Assuming $h(B,f)$ has a smooth dependance in magnetic field,

$$\frac{1}{h(B_0(f),f)} \times \left(\frac{\partial h}{\partial B}\right)_{(B_0(f),f)} \ll \frac{1}{\Gamma},$$

the measured peak position might be estimated as follow, shifted from the real resonance field $B_0(f)$

$$B_{measured}(f) \approx B_0(f) + \frac{\Gamma^2}{2a} \times \left(\frac{\partial h}{\partial B}\right)_{(B_0(f),f)}.$$

As expected, the lower is the amplitude of the peak, the larger is the effect of the baseline drift. Then, if one considers the second derivative signal $D_2$,

$$D_2 = \frac{\partial^2 S}{\partial B^2} = -\frac{2a}{\Gamma^2} \frac{1 - 3\left(\frac{B - B_0(f)}{\Gamma}\right)^2}{\left[1 + \left(\frac{B - B_0(f)}{\Gamma}\right)^2\right]^3} + \frac{\partial^2 h}{\partial B^2},$$

it is no longer affected by the quadratic dependance of the $h(B,f)$ since its second derivative will result in a constant shift of $D_2$ that will note impact the peak localization. Assuming a smooth dependance in magnetic field of the baseline signal, one expects that most of the baseline is mostly locally quadratic so that most of it is removed in the derivation process. Furthermore, the width of the

$D_2$ mean peak is smaller by a factor 3. The error in peak position scales as the square of the peak width and is proportional to the derivative of the baseline, as discussed previously. Then, one benefits both from a factor 9 in precision from the peak width and reduction of the amplitude of the baseline by removing its quadratic contribution.

Another quantity of interest is the crossed second derivative $D_{cross}$

$$D_{cross} = \frac{\partial^2 S}{\partial f \partial B} = 2a\Gamma^2 B_0'(f) \frac{\Gamma^2 - 3(B - B_0(f))^2}{\left(\Gamma^2 + (B - B_0(f))^2\right)^3},$$

$$\frac{\partial^3 S}{\partial B \partial f \partial B} = \frac{-24a\Gamma^2 B_0'(f)(B - B_0(f))\left(\Gamma^2 - (B - B_0(f))^2\right)}{\left(\Gamma^2 + (B - B_0(f))^2\right)^4}$$

in order to isolate only absorption lines that depends on the incident wave frequency. The quantity $D_{cross}$ will be extremal also for $B = B_0(f)$ but of non-zero value only if there is a dependence in frequency of the resonance magnetic field $B_0(f)$. This permits to suppress any pattern that is not dependent in frequency. Such a derivative is proportional to the line position in magnetic field derivative $B_0'(f)$, with three extrema values at $B_0(f)$ and $B_0(f) \pm \Gamma$ of opposite sign between the central ones and side ones. More precisely

$$\frac{\partial^2 S}{\partial f \partial B}\Big|_{B = B_0(f)} = 2a\Gamma^2 B_0'(f),$$

$$\frac{\partial^2 S}{\partial f \partial B}\Big|_{B = B_0(f) \pm \Gamma} = -\frac{a\Gamma^2}{2} B_0'(f),$$

so that sides peaks are four times lower in absolute amplitude. Then, lines will be localized on the absolute value of the $D_2$ signal (i.e. $|D_2|$) with a peak finding routine, so that only features with a frequency dependence are extracted. An appropriate threshold is defined so that side peaks of $|D_2|$ are not detected, since they have a lower amplitude. Absolute value is required as a result of signal distortion from long term drifts or noise increase from numerical derivation. In that aim, the *find_peak* method of *scipy.signal* python library has been used on each $D_2$ curve, as a function of $B$.

For a given frequency, $D_{cross}$ and $D_2$ are computed both for single layer and bilayer graphene sample. Peak of photoconductivity are automatically extracted from those data using *find_peak* routine of *scipy.signal* library in Python, with typically prominence parameter of 0.5, and peak distance of 100 mT. Then, these peaks are adjusted with a linear fit routine. On single layer graphene sample, one obtained the mapping in frequency of Fig. 1. Then, for each line, a region of interest is defined by hand from the ensemble of peaks detected, in order to provide linear regression fit procedure for each transition observed. The optimal parameters and their 95% confidence interval are estimated with the *linregress* method of the *scipy.stats* python library.

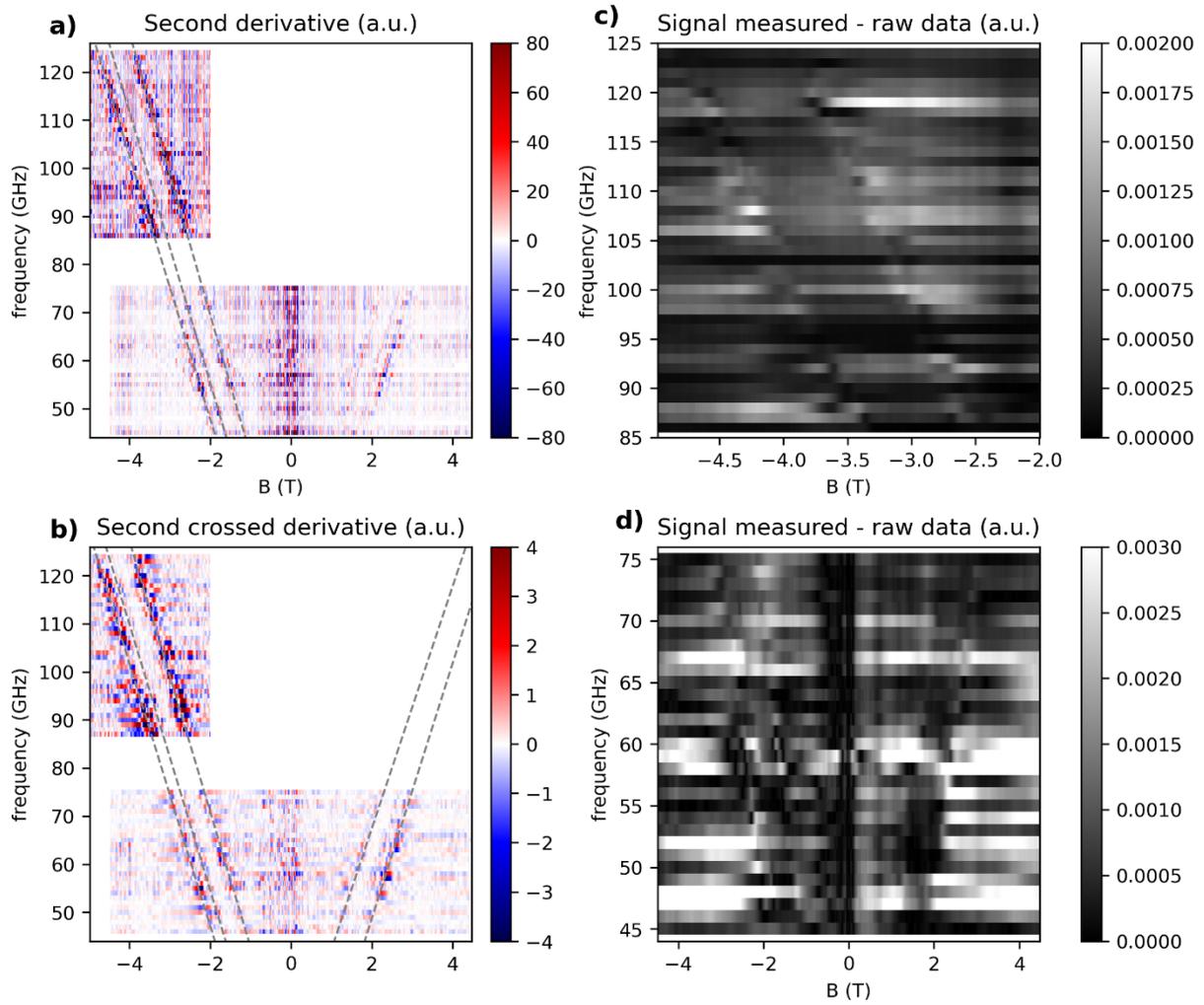

Figure 1 – a) Numerically calculated $D_2$ signal of single layer graphene sample. Dashed line corresponds to linear fit of automatically detected peaks. b) Numerically calculated $D_{cross}$ signal of single layer graphene sample. Dashed line corresponds to linear fit of automatically detected peaks. c) Raw data of photoconductive signal in the 85-125 GHz range. d) Raw data of photoconductive signal in the 45-75 GHz range.

From linear fits, one could extract the value of the slope and the frequency at zero-field (B=0T) for each transition observed. These values are summed up in table 1 for bilayer graphene (sample B) and in table 2 for single layer graphene (sample A). From these values, one may extract the value of the Landé g-factor, summarized in table 3.

| Line # | $\frac{\partial^2 S}{\partial B^2}$ | | $\frac{\partial^2 S}{\partial f \partial B}$ | |
|---|---|---|---|---|
| | Slope (GHz/T) | Frequency at B=0 (GHz) | Slope (GHz/T) | Frequency at B=0 (GHz) |

| Line # | ∂²S/∂B² Slope (GHz/T) | ∂²S/∂B² Frequency at B=0 (GHz) | ∂²S/∂f∂B Slope (GHz/T) | ∂²S/∂f∂B Frequency at B=0 (GHz) |
|---|---|---|---|---|
| 1 (γ') | $-27.4\pm0.3$ | $-7.0\pm0.9$ | $-27.3\pm0.3$ | $-7.4\pm1.5$ |
| 2 (α) | $-28.2\pm0.4$ | $-1.1\pm1.0$ | $-27.6\pm0.8$ | $-0.7\pm2.3$ |
| 3 (β) | $-29.3\pm0.4$ | $11.0\pm1.0$ | $-28.1\pm0.4$ | $14.6\pm1.7$ |
| 4 (β) | - | - | $25.2\pm1.6$ | $17.1\pm2.9$ |
| 5 (α) | - | - | - | - |
| 6 (γ') | - | - | $26.5\pm0.9$ | $-4.4\pm2.6$ |

Table 1 – Value of slope and frequency at zero field for each transition observed in the case of a bilayer graphene sample. Values are obtained from a linear fit of the detected peak of second derivative signal.

| Line # | $\frac{\partial^2 S}{\partial B^2}$ Slope (GHz/T) | $\frac{\partial^2 S}{\partial B^2}$ Frequency at B=0 (GHz) | $\frac{\partial^2 S}{\partial f \partial B}$ Slope (GHz/T) | $\frac{\partial^2 S}{\partial f \partial B}$ Frequency at B=0 (GHz) |
|---|---|---|---|---|
| 1 (γ') | $-27.7\pm0.4$ | $-6.2\pm1.5$ | $-27.5\pm0.4$ | $-5.7\pm1.9$ |
| 2 (α) | $-28.0\pm0.4$ | $-0.4\pm1.4$ | $-28.7\pm0.4$ | $-3\pm1.9$ |
| 3 (β) | $-28.35\pm0.5$ | $12.5\pm1.6$ | $-29.3\pm0.8$ | $9.2\pm3.2$ |
| 4 (β) | $29.9\pm1.0$ | $7.8\pm3.4$ | $28.8\pm0.7$ | $11.0\pm2.7$ |
| 5 (α) | - | - | $28.3\pm0.61$ | $-0.9\pm2.6$ |
| 6 (γ') | $27.2\pm0.5$ | $-3.6\pm2.0$ | $27.9\pm0.6$ | $-6.1\pm2.7$ |

Table 2 – Value of slope and frequency at zero field for each transition observed in the case of a single layer graphene sample. Values are obtained from a linear fit of the detected peak of second derivative signal.

| | Single Layer Graphene | | Bilayer Graphene | |
|---|---|---|---|---|
| Line # | $\frac{\partial^2 S}{\partial B^2}$ | $\frac{\partial^2 S}{\partial f \partial B}$ | $\frac{\partial^2 S}{\partial B^2}$ | $\frac{\partial^2 S}{\partial f \partial B}$ |
| 1 (γ') | $1.96 \pm 0.02$ | $1.95 \pm 0.02$ | $1.98 \pm 0.03$ | $1.96 \pm 0.03$ |
| 2 (α) | $2.01 \pm 0.03$ | $1.97 \pm 0.06$ | $2 \pm 0.03$ | $2.05 \pm 0.03$ |
| 3 (β) | $2.09 \pm 0.03$ | $2.01 \pm 0.03$ | $2.025 \pm 0.04$ | $2.09 \pm 0.06$ |
| 4 (β) | - | $1.8 \pm 0.11$ | $2.14 \pm 0.07$ | $2.06 \pm 0.05$ |
| 5 (α) | - | - | - | $2.02 \pm 0.04$ |
| 6 (γ') | - | $1.89 \pm 0.06$ | $1.94 \pm 0.04$ | $1.99 \pm 0.04$ |
| Mean value | $2.02 \pm 0.02$ | $1.92 \pm 0.06$ | $2.02 \pm 0.02$ | $2.03 \pm 0.02$ |

Table 3 – Estimation of Landé g-factor for single and bilayer graphene sample.

| Line # | Bilayer graphene | | | | Single layer graphene | | | |
|---|---|---|---|---|---|---|---|---|
| | $\frac{\partial^2 S}{\partial B^2}$ | | $\frac{\partial^2 S}{\partial f \partial B}$ | | $\frac{\partial^2 S}{\partial B^2}$ | | $\frac{\partial^2 S}{\partial f \partial B}$ | |
| | Slope (GHz/T) | Frequency at B=0 (GHz) | Slope (GHz/T) | Frequency at B=0 (GHz) | Slope (GHz/T) | Frequency at B=0 (GHz) | Slope (GHz/T) | Frequency at B=0 (GHz) |
| 1 (γ') | − 27.7±0.4 | − 6.2±1.5 | − 27.5±0.4 | − 5.7±1.9 | − 27.4±0.3 | − 7.0±0.9 | − 27.3±0.3 | − 7.4±1.5 |
| 2 (α) | − 28.0±0.4 | − 0.4±1.4 | − 28.7±0.4 | − 3±1.9 | − 28.2±0.4 | − 1.1±1.0 | − 27.6±0.8 | − 0.7±2.3 |
| 3 (β) | − 28.35±0.5 | 12.5±1.6 | − 29.3±0.8 | 9.2±3.2 | − 29.3±0.4 | 11.0±1.0 | − 28.1±0.4 | 14.6±1.7 |
| 4 (β) | 29.9±1.0 | 7.8±3.4 | 28.8±0.7 | 11.0±2.7 | - | - | 25.2±1.6 | 17.1±2.9 |
| 5 (α) | - | - | 28.3±0.61 | − 0.9±2.6 | - | - | - | - |
| 6 (γ') | 27.2±0.5 | − 3.6±2.0 | 27.9±0.6 | − 6.1±2.7 | - | - | 26.5±0.9 | − 4.4±2.6 |

Table 4 – Estimated slopes and zero field frequencies by two different methods for single layer and bilayer graphene.

**Temperature dependence of peak position**

The temperature dependences of observed line have been also studied in the case of bilayer graphene, summed up in Fig. 2. These results are consistent with observations made in single layer graphene.

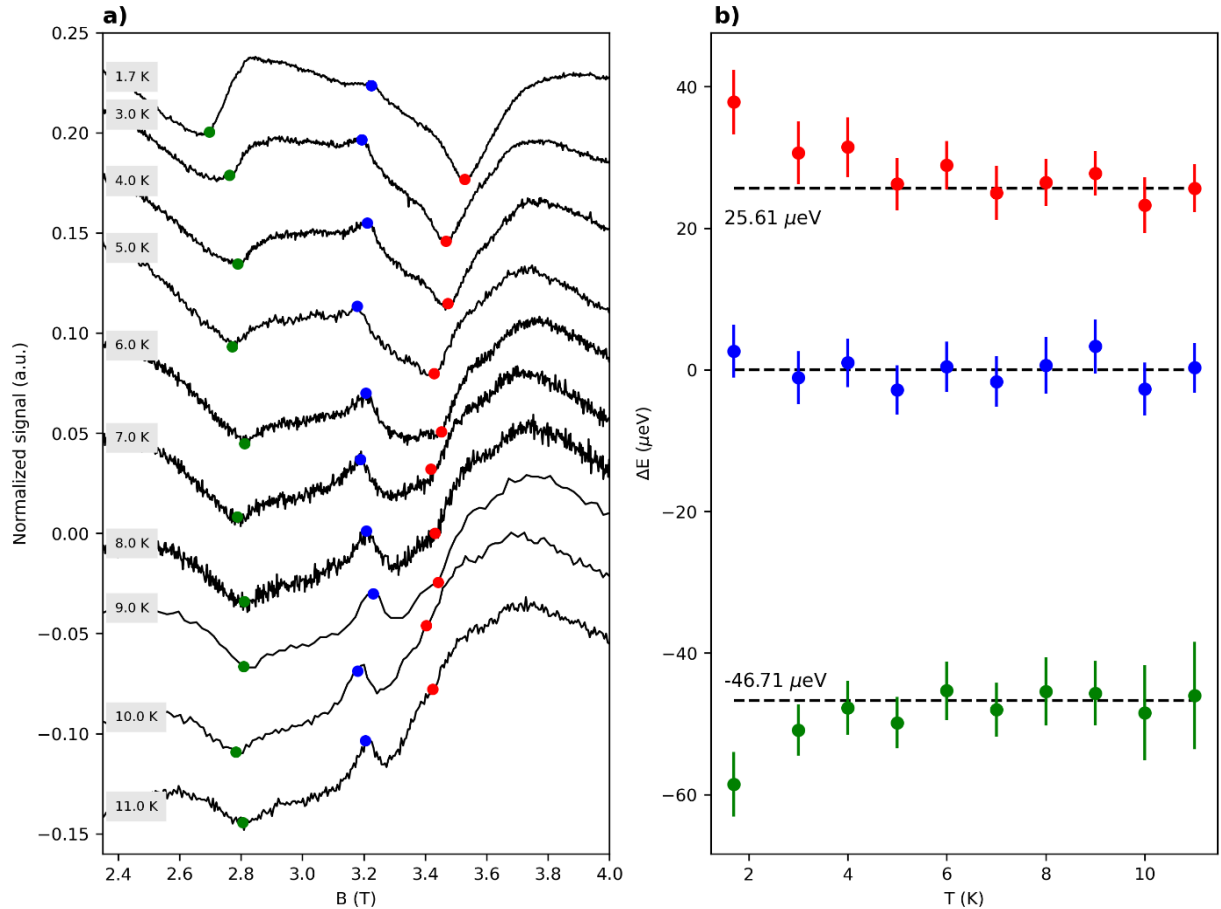

Figure 2 - Evolution of the resonance in bilayer graphene versus the temperature for a source frequency of 90 GHz. a) Photoresponse is plotted versus magnetic field at different temperatures. Red, blue and green dots represent the evolution with the temperature for the α, β and γ' transitions respectively. No effect of temperature is observed for the zero-field valley transition as opposed as the intrinsic spin-orbit coupling splitting ones. b) Energy of the intercept at zero origin versus the temperature for the three observed transitions. Same color as a) is used to represent the transition. Dash line represents the converge values of energy at high temperature, -47 and 26 μeV for β and γ' transitions respectively.

Besides this temperature-induced strain effect, the spin resonance shift could indicate the presence of low-temperature magnetic ordering [1]. The analysis of the amplitude of the resonances, proportional to the paramagnetic susceptibility, makes it possible to separate a ferromagnetic signal from a

---

paramagnetic contribution [2]. It is impossible to retrace the evolution of the intensity of the α line because it is too weak in all of our results. However, we plot the amplitude of the second derivative of the signal for the γ' and β transitions over the entire temperature range in Fig. 3, measured in sample B at a frequency of 60 GHz. The amplitude of the second derivative is proportional to the amplitude of the resonance of the lock-in signal, as explained in previous section "Data processing". Even though this is a preliminary study, it is clear that the signal obeys a Curie law with a Curie-Weiss temperature tending to 0, which indicates the absence of any magnetic order.

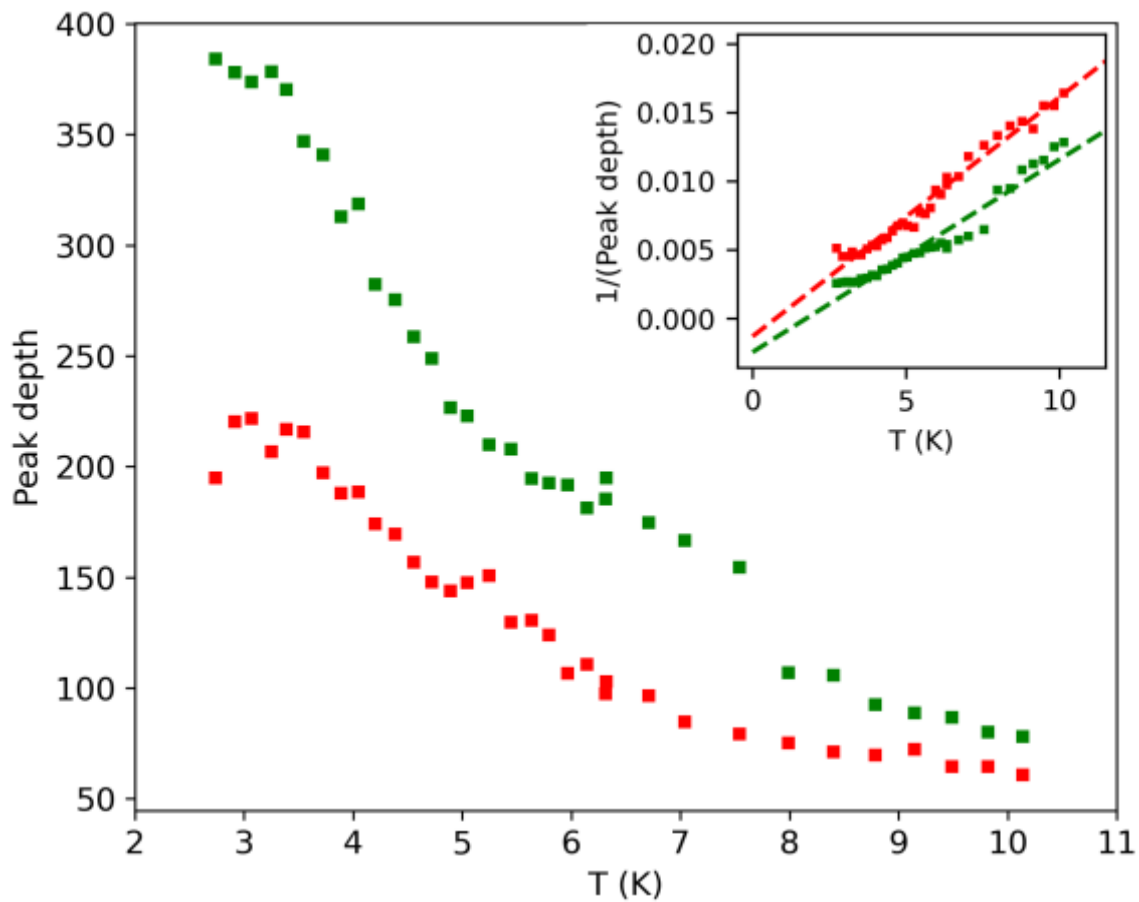

Figure 3 - Evolution of the amplitude of the optical transitions plotted versus temperature for sample A at 60 GHz. Red and green dots represent γ' and β transitions, respectively. Insert shows the reverse of the amplitude of the optical transitions. Dashed lines correspond to linear fits.

**Temperature dependence of peak amplitude**

Since the probe frequency (60GHz) involves rather low energies of 240µeV (3 K equivalent temperature), one expects thermal activation to play a significant role in the temperature dependence of signal peak depth. Considering a two-level model, with a zero-field splitting of $\Delta$, population of upper and lower level are given by Fermi-Dirac distribution $f(\varepsilon, \mu) = \frac{1}{1+e^{\beta(\varepsilon-\mu)}}$, with $\beta = \frac{1}{k_B T}$ and $\mu$ the chemical potential. For a given magnetic field $B$, levels energies might be written as follow

$$E_1 = -\frac{\Delta}{2} - \mu_B g B \quad and \quad E_2 = \frac{\Delta}{2} + \mu_B g B,$$

with $g$ the Landé factor. For $\nu$ the probe frequency ($\nu = 60\text{GHz}$), the resonance condition of the transition observed associated to a given peak provide the following relationship

$$h\nu = E_2 - E_1 = \Delta + 2\mu_B g B.$$

The probability $P(\Delta, \beta, \mu, \nu)$ for absorbing a probe photon is proportional to the probability of having an electron in the lower level and no electron in the upper level is then given by

$$P(\Delta, \beta, \mu) = f(E_1, \mu) \times \left(1 - f(E_2, \mu)\right) = \frac{1}{1+2\cosh\cosh\,(\beta\mu)\,e^{-\beta\left(\frac{\Delta}{2}+\mu_B g B\right)} + e^{-\beta(\Delta+2\mu_B g B)}}.$$

Then, introducing the probe frequency $\nu$, it might be rewritten as follow

$$P(\beta, \mu, \nu) = \frac{1}{1+2\cosh\cosh\,(\beta\mu)\,e^{-\frac{\beta h\nu}{2}} + e^{-\beta h\nu}}.$$

The change in conductivity related to resonant absorption of the incident probe wave is proportional to the probability $P(\Delta, \beta, \mu, \nu)$, with a coefficient denoted $a$ that depends on the experimental setup. Moreover, the detection scheme of this work is based on Ratchet devices that are non-linear process. Consequently, as a nonlinear process, a threshold is expected in the signal generation. Additionally, regarding level of noise in the experimental data, and the numerical method used for peak amplitude measurements, an additional offset $b$ in the model is added. Such simple aims at confirming that temperature dependence of peaks amplitude is dominated by thermal activation effects, using realistic parameters, and modelling technical offsets from reasonable assumptions.

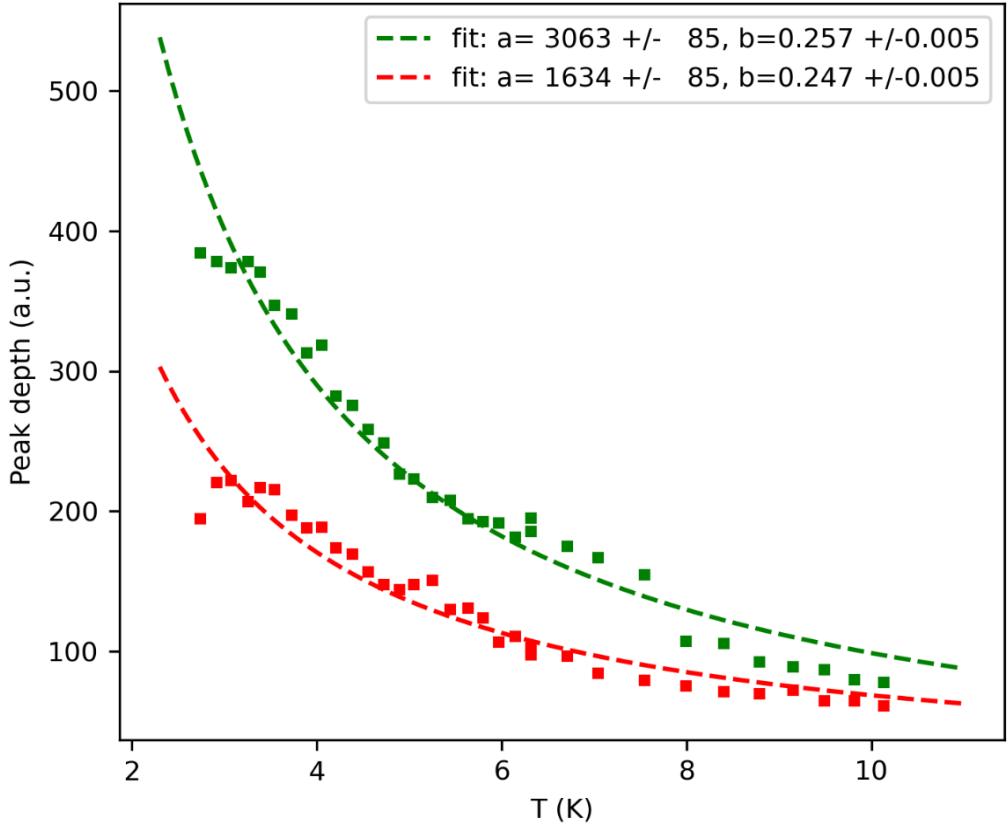

Figure 4 - Evolution of the amplitude of the optical transitions plotted versus temperature for sample A at 60 GHz. Red and green dots represent γ' and β transitions, respectively. Dashed line corresponds the theoretical model fitted data, with resulting fit parameters in legend.

Then one may propose the following analytical expression of the detected signal $S(\Delta, \beta, \mu, \nu)$ amplitude at resonance

$$S(\beta, \mu, \nu) = a \times \left( \frac{1}{1 + 2\cosh(\beta\mu)\, e^{-\frac{\beta h\nu}{2}} + e^{-\beta h\nu}} - b \right),$$

where $a$ and $b$ are unknown parameters related to the experimental setup, that might be taken as fit parameters. from reasonable assumptions. This analytical model has been fitted with amplitude of γ' and β lines measured at 60 GHz for sample A (same data as previous figure, Fig. 3). Data have been fitted using curve_fit routine from Python 3.0 scipy.optimize library, providing good agreement between data and model as illustrated in Fig. 4. Results of fits are providing the same offset coefficient within confidence intervals obtained. Within confidence intervals of fitting parameters $a$, γ' line strength is twice the strength of β line.

**Measurements on THz detector based on the graphene p-n junction**

The graphene detector made from a p-n junction was measured in the same way as the ratchet detectors. It was placed in a 6 T horizontal cryogen free magnet system with optical access. The voltage generator (Keitley 2600B) allowed to control the current applied on the device. The measurements were done at 1.7 K by applying a current of -140 µA on the device. Sub-THz source generated by Shottky diode of multiplied frequency were used to obtain frequencies in the ranges from 82 GHz to 125 GHz (optical power about 10 mW). The signal was detected as a voltage drop at the edges of the detector, which is then amplified and measured via a standard lock-in technique (using an Amatek Signal recovery 6270).

Although the intensity of the photoconductivity signal and the signal-to-noise ratio obtained in the p-n junction were lower than with the ratchet devices, we were able to detect spin resonances and followed their evolution as a function of the incident frequency and the magnetic field. It turns out that the positions of the resonances are comparable to those obtained in the ratchet detectors. The results obtained with this p-n junction-based sensor as a function of the incident frequency are plotted in the form of a color mapping in Figure 5. For each line, a region of interest is manually defined from the set of detected peaks, in order to provide a linear fitting procedure for each observed transition. The optimal parameters and their 95% confidence interval are estimated with the linregress method of the python scipy.stats library. From the linear fits, it was possible to extract the value of the slope and the zero field frequency (B = 0 T) for each observed transition. These values are summarized in Table 5.

Two spin resonances in the p-n junction of graphene are observed, whose zero-field extrapolation allows their identification. The first gives a g-factor roughly equal to two, just like that measured in ratchet detectors. The second allows us to extract a zero-field splitting of the order of 17.6 GHz (73 µeV), instead of the approximately 13 GHz observed in ratchet devices at low temperature. It should be noted that the measurement uncertainty is high given the low signal-to-noise ratio. We also underline that the geometry of the p-n junction detector requires a wide log-periodic metallic antenna placed on the graphene sheet to effectively rectify the incident THz wave. The strain effects on the graphene layer should therefore be significant at low temperature and could be responsible for this higher ZFS value. Unfortunately, it was impossible to follow the evolution of this ZFS value as a function of temperature because the intensity of the signal dropped below the detection threshold at temperatures slightly higher than that used for these results.

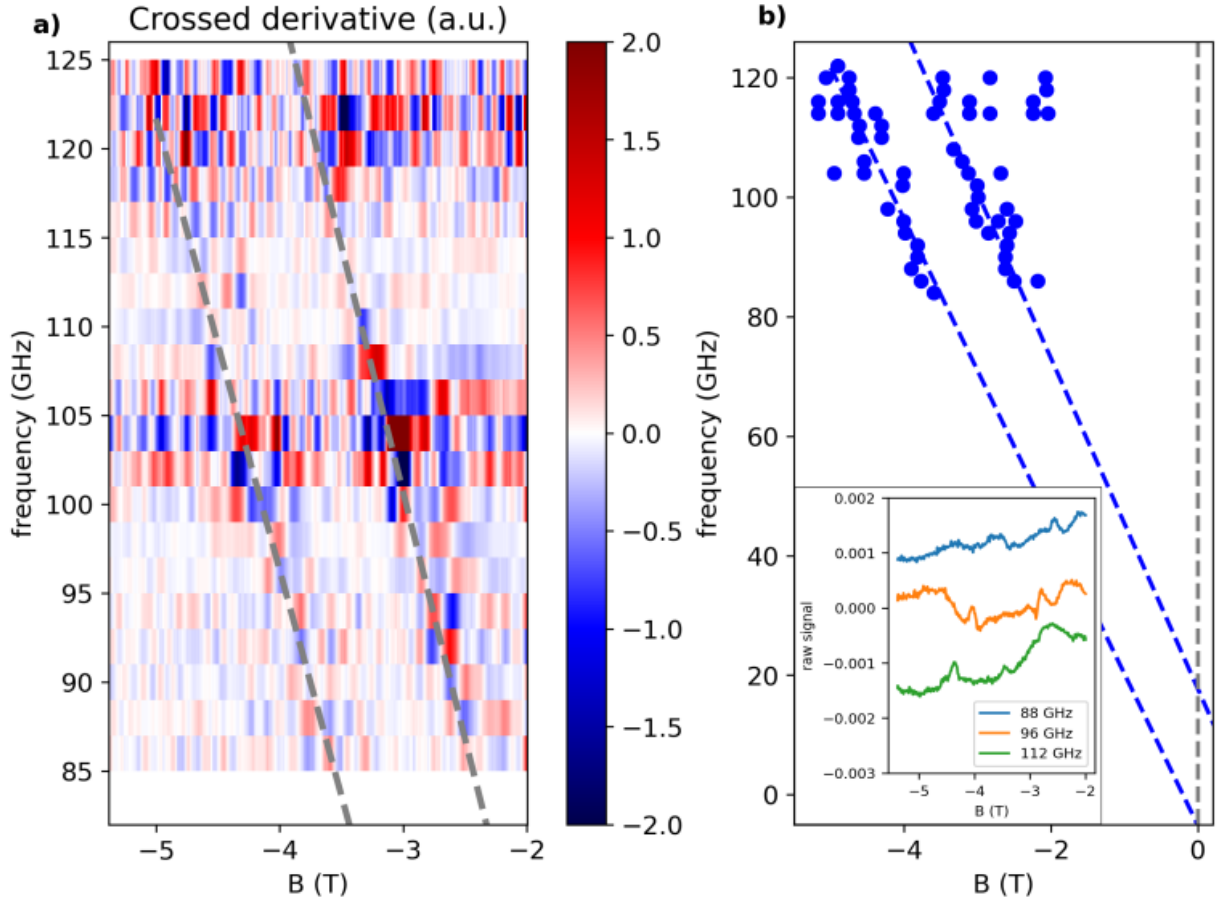

Figure 5 – a) Numerically calculated $D_{cross}$ signal of graphene detector with a p-n junction. Dashed line corresponds to linear fit of automatically detected peaks. b) Frequency of the different resonances as a function of magnetic field in the range 82-125 GHz measured at 1.7 K. The dashed lines are some linear fits obtain from measurement in the 82-125GHz range allowing for the extrapolation of the experimental results to B = 0. The lower dashed line follows the linear evolution with a slope of -25.4 GHz/T ± 1.5 GHz/T and with an intercept frequency at B =0 of -5.2 ± 6.6 GHz. The other dashed lines have a slope of -27.7 GHz/T ± 1.6 GHz/T and with an intercept frequency at B = 0 of 17.6 ±4.8 GHz. The inset is example of raw data of photoconductive signal in the 85-125GHz range.

| Line # | $\frac{\partial^2 S}{\partial f \partial B}$ | | | |
|--------|-----------------|-----------------|------------------------|-----------------|
|        | Slope (GHz/T)   | Landé g-factor  | Frequency at B=0 (GHz) | ZFS (µeV)       |
| 1      | $-25.4\pm1.5$   | $1.8\pm0.11$    | $-5.2\pm6.6$           | $-21.6\pm27.2$  |
| 2      | $-27.7\pm1.6$   | $2.0\pm0.11$    | $17.6\pm4.8$           | $73.0\pm19.7$   |

Table 5 – Estimated slopes and zero field frequencies for graphene detector with a p-n junction.